\documentclass[twocolumn]{aastex631}

\begin{document}

\title{{\em Suzaku} Observations of the Cold Front in A3667: a Non-Equilibrium Ionization State of the Intracluster Medium}

\author[0009-0002-7226-5019]{Mona Shishido}
\affiliation{Department of Physics and Astronomy, Tokyo University of Science, 2641 Yamazaki, Noda, Chiba 278-8510, Japan}
\affiliation{Academia Sinica Institute of Astronomy and Astrophysics (ASIAA), 11F of AS/NTU Astronomy-Mathematics Building, No.1, Sec. 4, Roosevelt Rd, Taipei 106216, Taiwan}

\author[0000-0002-7962-4136]{Yuusuke Uchida}
\affiliation{Department of Physics and Astronomy, Tokyo University of Science, 2641 Yamazaki, Noda, Chiba 278-8510, Japan}


\author[0000-0003-4403-4512]{Takayoshi Kohmura}
\affiliation{Department of Physics and Astronomy, Tokyo University of Science, 2641 Yamazaki, Noda, Chiba 278-8510, Japan}

\begin{abstract}
We present a systematic study of the thermodynamic properties of the intracluster medium (ICM), including the plasma conditions in the ICM, in the merging galaxy cluster A3667 using {\em Suzaku}. We analyze the X-ray spectra of the ICM between the northwestern and the southeastern radio relics across a prominent cold front in A3667 to search for the ICM in a non-equilibrium ionization (NEI) state. We find that the ICM inside the cold front exhibits an NEI state with an ionization parameter of $(2.5\pm0.7)\times10^{11}\,\mathrm{s\,cm^{-3}}$ at the $90\%$ confidence level, which is lower than that expected from a collisional ionization equilibrium (CIE) state of the ICM (i.e., $n_\mathrm{e}t > 10^{12}\,\mathrm{s\,cm^{-3}}$). The timescale calculated from the ionization parameter of the ICM inside the cold front is $ 4.6\pm1.2\,\mathrm{Myr}$, which is much shorter than the thermal equilibration timescale. A weak transonic sloshing motion might explain the possibility that the ICM inside the cold front of A3667 is in the NEI state. In addition, the NEI state of the ICM in A3667 seems to be associated with the elongated radio halo bridging the region between the northwestern radio relic and the prominent cold front in A3667 across the cluster center.  
\end{abstract}


\section{Introduction}
\label{sec:intro}

Galaxy clusters are the largest and most massive gravitationally bound objects in the Universe. They evolve through mergers with smaller or similar-mass galaxy clusters. Dynamical processes induced by cluster mergers play an important role in shaping the structures and properties of galaxy clusters. In particular, cluster mergers drastically change the dynamic and thermodynamic properties of the diffuse, hot, X-ray emitting gas known as the intracluster medium (ICM), which has a temperature of $10^{7} - 10^{8}$\,K. The ICM represents the largest fraction of baryons trapped in the deep gravitational potential well of dark matter halos and thus serves as a probe for exploring high-energy astrophysical phenomena associated with cluster mergers.

One of the most important phenomena resulting from cluster mergers is shocks \citep{Markevitch2007}. Shocks are expected to heat the ICM and induce turbulent motions of the ICM, leading to huge impacts on its nature. X-ray observations with high-angular resolution have detected shock fronts in the X-ray surface brightness of merging galaxy clusters \citep[e.g.,][]{Markevitch2002, Markevitch2005, Russell2012, Rahaman2022}. In addition, cold fronts are thought to be formed by cluster mergers \citep[e.g.,][]{Vikhlinin2001, Vikhlinin2001a, Tittley2005, Markevitch2007, Owers2009, Ueda2017, Botteon2018}. Cold fronts are located at the interface between cooler, denser gas (i.e., lower entropy gas) and hotter, less dense gas (i.e., higher entropy gas) \citep[e.g.,][]{Vikhlinin2001, Markevitch2007, ZuHone2016}. Since cold fronts exhibit sharp edges in the X-ray surface brightness, not only ram-pressure but also the microphysical properties of the ICM provide non-thermal support for maintaining the sharpness of cold fronts \citep[e.g.,][]{Markevitch2007, Ichinohe2019}. Both shock fronts and cold fronts are important for understanding the impact of cluster mergers on the dynamic and thermodynamic properties of the ICM.

Diffuse synchrotron emission at radio wavelengths has often been found in the outskirts of galaxy clusters \citep[e.g.][]{Giovannini1991, Ferrari2008, Paul2020, Duchesne2021, Chibueze2023, Koribalski2024}. Such diffuse radio emissions are known as radio relics and are thought to be caused by shocks with $\sim\mu\mathrm{G}$ magnetic fields and relativistic particles. Radio relics have been reported to be associated with shock fronts seen in the X-ray surface brightness \citep[e.g.,][]{Giacintucci2008, Akamatsu2011, Eckert2016, Urdampilleta2018, Zhang2020, Botteon2020}. On the other hand, radio halos are also commonly found \citep[e.g.,][]{Willson1970, Large1959, Giovannini2009, Bonafede2012, Wilber2018, Wilber2020}. Radio halos are typically 1\,Mpc in size and are thought to be formed by cluster mergers. The presence of radio halos indicates ongoing merging activities. However, in contrast to radio relics, radio halos show no corresponding features in the X-ray surface brightness of galaxy clusters.

It has been predicted that shocks induce deviations from a collisional ionization equilibrium (CIE) in the ICM, leading to a non-equilibrium ionization (NEI) state of the ICM \citep{Masai1984, Takizawa1999, Akahori2010}. However, it is challenging to observe the NEI state because its survival timescale is very short (typically less than $\sim 10^{7}$\,yr). Despite this difficulty, searches for the NEI state of the ICM have been conducted \citep[e.g.,][]{Fujita2008, Russell2012, Inoue2016}. \cite{Inoue2016} found the NEI state of the ICM located at the shock front in A754, one of the major merging clusters, with {\em Suzaku} for the first time. The presence of the NEI state of the ICM is important for investigating energy equilibrium between electrons and protons \citep{Markevitch2006}.

A3667 is a nearby, X-ray bright galaxy cluster at $z=0.0556$ \citep{Struble1999}, and has the mean temperature of $7.0\pm0.6\,\mathrm{keV}$ \citep{Markevitch1998, Vikhlinin2001a, Reiprich2002}. The pair of bright giant radio relics of A3667 has been vigorously investigated \citep[e.g.,][]{Rottgering1997, Carretti2013, Hindson2014, Riseley2015, Gasperin2022}, and the cold front of A3667 has been well studied \citep[e.g.,][]{Markevitch1999, Vikhlinin2001a, Ichinohe2017, Omiya2024, Ueda2024}. In addition, \cite{Gasperin2022} presented an elongated radio halo running from the northwest to the cold front with MeerKAT.

\cite{Nakazawa2009} reported the presence of a very hot gas with a temperature of $19.2_{-4.0}^{+4.7}~\mathrm{keV}$ using {\em Suzaku}. \cite{Akamatsu2012} and \cite{Akamatsu2013} studied the shock fronts with {\em Suzaku} and discussed the X-ray and radio properties of the ICM associated with the shock fronts. However, although {\em Suzaku} observations have been conducted to cover the entire region between the northwestern and southeastern radio relics, no systematic study of A3667, such as the global properties of the ICM, has been carried out. In particular, there has been no detailed study of the cold front in A3667 using {\em Suzaku}. Since all products, including the final version of the calibration database (CALDB), are now available for {\em Suzaku}, we perform a systematic study of A3667 using {\em Suzaku} to reveal the nature of the ICM affected by the cluster merger activities in this paper. The gain uncertainty of the X-ray CCD camera on board {\em Suzaku} has been well calibrated \citep[e.g.,][]{Uchiyama2009, Tamura2011}, such that {\em Suzaku} observations provide us with an opportunity to probe the ionization state of the ICM, as demonstrated by \cite{Inoue2016}.

This paper is organized as follows. Section~\ref{sec:obs} describes the {\em Suzaku} observations of A3667 and data reduction. Section~\ref{sec:ana} describes the X-ray data analysis for A3667. In Section~\ref{sec:discussion}, we discuss the obtained results and their implications. Finally, the conclusions of this paper are summarized in Section~\ref{sec:conclusions}.

We assume a spatially flat Lambda Cold Dark Matter ($\Lambda$CDM) model with $H_0=70\,\mathrm{km\,s^{-1}\,Mpc^{-1}}, \Omega_\mathrm{M}=0.27, \Omega_\Lambda=1-\Omega_\mathrm{M}=0.73$. In this cosmology, an angular size of $1'$ corresponds to a physical scale of 66\,kpc at $z=0.0556$. All errors are given at 90$\%$ confidence level unless stated.

\section{Observations and Data Reduction} \label{sec:obs}

\subsection{Observation Logs}

We analyze the archival X-ray datasets of A3667 taken with the X-ray Imaging Spectrometer \citep[XIS;][]{Koyama2007} on board {\em Suzaku}. Additionally, we use the {\em Chandra} archival datasets for A3667 obtained with ACIS-I \citep[i.e., the Advanced CCD Imaging Spectrometer (ACIS) with the front-illuminated chip array;][]{Garmire2003, Grant2024}. Table~\ref{tbl:observation_logs} summarizes the observation logs of A3667 from both {\em Suzaku} and {\em Chandra} observations.

\begin{figure}[htbp]
   \begin{center}
      \includegraphics[width=8.5cm, trim={4.2cm 8.69cm 4.2cm 7.5cm},clip]{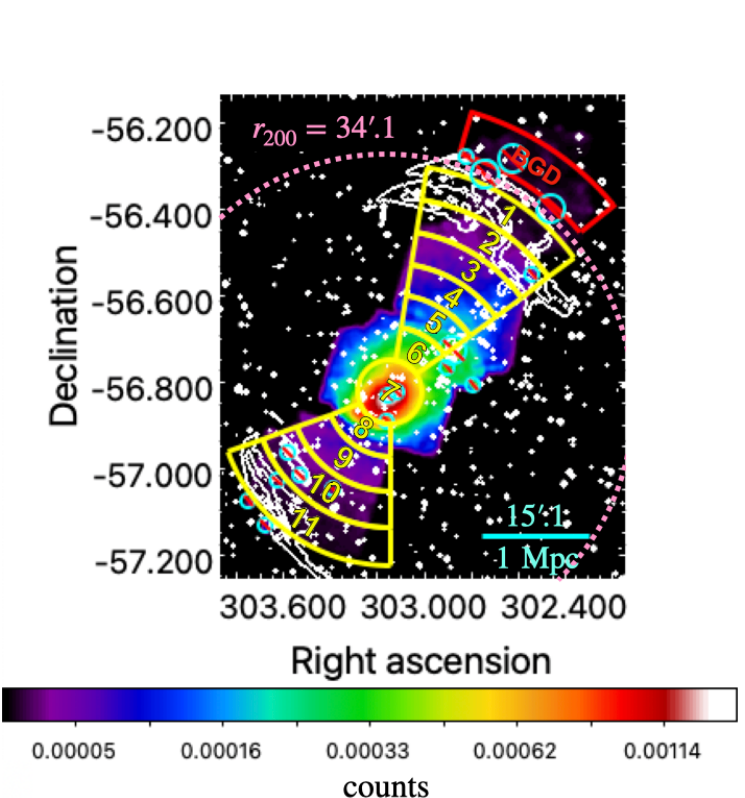}
   \end{center}
   \caption{X-ray surface brightness image of A3667 taken with {\em Suzaku} FI-CCDs (i.e., XIS0, XIS2, and XIS3) in the $0.7-7.0\,\mathrm{keV}$ band. The NXB contribution is subtracted, and the vignetting correction is applied. This image is smoothed by a Gaussian kernel with a width of $4.7$\,arcsec. The yellow sectors show the regions used to conduct the X-ray spectral analysis. The red sector represents the region used to estimate the sky X-ray background. The cyan circles show the point sources detected by {\em XMM-Newton}. The white contours show the diffuse synchrotron emission at $1.4\,\mathrm{GHz}$ observed with MeerKAT.
    }\label{fig:suzaku_image}
\end{figure}

\begin{deluxetable*}{cccccc }[htb]
    \tablecaption{Summary of the {\em Suzaku} and {\em Chandra} Observation logs of A3667.} \label{tbl:observation_logs}
    \tablehead{
    \colhead{Observatory} & \colhead{Object Name} & \colhead{ObsID} & \colhead{Date-Obs} & \colhead{( RA, Dec )} & \colhead{Exposure [ks]} }
    \startdata
        {\em Suzaku}&NW$\_$RELIC & 801094010&  2006-05-03T17:47:01&	( 302.591, -56.413 )&94.4\\
        &17OFF1 & 801095010&  2006-05-06T07:03:14&	( 302.848, -56.613 )&17.3\\
        &17OFF2 & 801095020&  2006-10-30T19:58:53&	( 302.848, -56.613 )&12.0\\
        &CENTER & 801096010&  2006-05-06T17:40:17&	( 303.171, -56.850 )&20.9\\
        &SE$\_$RELIC & 805036010&  2010-04-12T04:00:59&	( 303.465, -57.031 )&66.3\\
        \hline
        {\em Chandra}&A3667& 513 &  1999-09-22T21:56:21&	( 303.210, -56.849 )&44.8\\
        &A3667& 889 &  2000-09-09T13:50:48&	( 302.958, -56.759 )&50.3\\
        &A3667& 5751 &  2005-06-07T00:46:21&	( 303.280, -56.890 )&128.9\\
        &A3667& 5752 &  2005-06-12T06:37:57&	( 303.280, -56.890 )&60.4\\
        &A3667& 5753 &  2005-06-17T04:55:02&	( 303.280, -56.890 )&103.6\\
        &A3667& 6292 &  2005-06-10T15:08:12&	( 303.280, -56.890 )&46.7\\
        &A3667& 6295 &  2005-06-15T06:02:17&	( 303.280, -56.890 )&49.5\\
        &A3667& 6296 &  2005-06-19T19:47:15&	( 303.280, -56.890 )&49.4\\
    \enddata
\end{deluxetable*}

\subsection{Data Reduction}
The XIS consists of 4 CCD chips (XIS0, XIS1, XIS2, and XIS3). XIS1 is a back-illuminated (BI) CCD chip, while XIS0, XIS2, and XIS3 are front-illuminated (FI) CCD chips. We use all the data of A3667 taken with the XIS in this paper. Note that XIS2 was no longer operating after November 2006 due to significant charge leakage. Therefore, for the 2010 observations (OBSID: 805036010), we use the datasets taken with XIS0 and XIS3. For the observations before November 2006 (OBSID: 801094010, 801095010, 801095020, 801096010), we combine all the datasets taken with XIS0, XIS2, and XIS3. All observations for A3667 have been performed with either the normal $3\times 3$ or $5\times 5$ clocking mode, and the datasets obtained with the $3\times 3$ or $5\times 5$ editing modes for each pointing are combined.

We use HEAsoft version 6.31.1 \citep{NHEASARC2014}, $\tt{XSPEC}$ version 12.13.0 \citep{Arnaud1996}, and the corresponding calibration database. We reprocess all the XIS datasets for A3667 using the $\tt{aepipeline}$ task in HEAsoft and select good time intervals in the satellite's orbit. We exclude the time intervals when the satellite passes through the South Atlantic Anomaly ($\mathrm{SAA (T\_SAA\_HXD)} > 436$) and the geomagnetic cut-off rigidity (COR) being above $6\,\mathrm{GV}$. Additionally, we limit the elevation angle (ELV) from the Earth rim to be above $10^\circ$ to avoid contamination from scattered solar X-rays from the day Earth limb. We use the $\tt{XSELECT}$ task in HEAsoft to extract X-ray spectra of the ICM from the standard filtered event files. Response matrix files (RMFs) and auxiliary response files (ARFs) are generated by the $\tt{xisrmfgen}$ task and the $\tt{xissimarfgen}$ task \citep{Ishisaki2007} in HEAsoft, respectively. To estimate the sky X-ray background contributions, we generate ARFs assuming a uniform emission distribution with a radius of $20'$.

\section{Analysis and results}
\label{sec:ana}

\subsection{Estimate of the sky X-ray background components}
\label{sec:bkg}

\subsubsection{Point Sources}
\label{sec:point}

Identifying and masking point sources is important to analyze faint, diffuse X-ray emission from the ICM, particularly in the outskirts. We first search for point sources in the field-of-view (FOV) of the {\em Suzaku} observations. We adopt the X-ray flux of $1 \times 10^{-13}\,\mathrm{erg\,cm^{-2}\,s^{-1}}$ as the threshold for point source detection. We detect 14 point sources coincident with the 3XMM-DR4 catalogue \citep{Watson2009}, which is based on {\em XMM-Newton} observations with better spatial resolution than {\em Suzaku}. We also identify 3 bright point sources near the northwestern relic region from the X-ray surface brightness image of A3667 taken with {\em Suzaku}. We set a radius of $1'$ or $2'$ to mask point sources, corresponding to a half-power diameter (HPD) of the X-ray telescopes \citep[XRT:][]{Serlemitsos2007} on board {\em Suzaku}. The point sources masked are shown in Figure \ref{fig:suzaku_image}.

\subsubsection{Sky X-ray background}
\label{sec:cxb}

\begin{figure*}[htb]
         \centering
         \includegraphics[width=18.1cm]{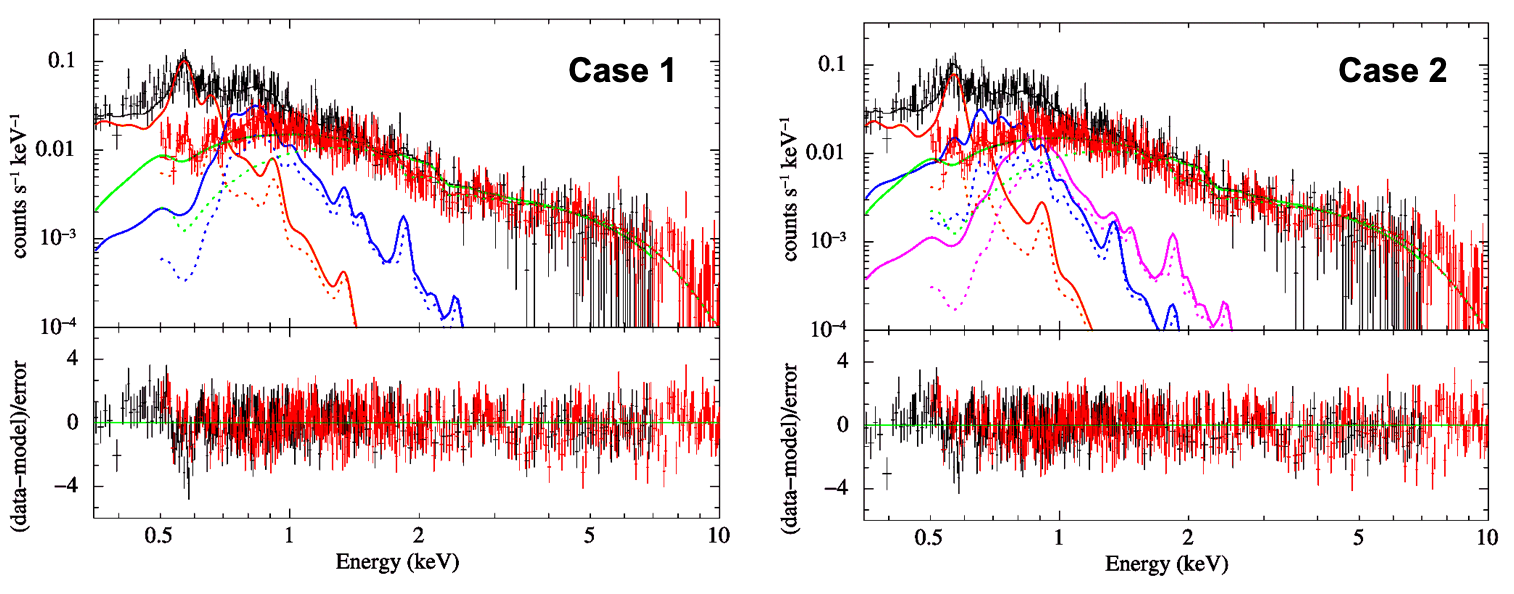}
   \caption{
   X-ray spectra extracted from the region out of $r_\mathrm{200} $ of A3667 with {\em Suzaku} (i.e., the BGD region in Figure~\ref{fig:suzaku_image}). The NXB contribution is subtracted. The black data points represent the count rates in each energy band observed with XIS1, and the red data points show the average count rate observed among XIS0, XIS2, and XIS3. The solid lines and the dotted lines show the best-fit models for XIS0 and XIS1, XIS2, and XIS3, respectively. The LHB, MWH, CXB, and thermal plasma emission with a temperature of $0.8\,\mathrm{keV}$ models are shown in the orange, blue, green, and magenta lines, respectively. \textit{Left}: the observed X-ray spectra and their best-fit models fitted with Case~1. \textit{Right}: Same as \textit{Left}, but for those fitted with Case~2.}
   \label{fig:result_bgd}
\end{figure*}

For the analysis of the X-ray emission from the ICM, it is important to evaluate accurately the sky X-ray background components, i.e., the cosmic X-ray background (CXB) and the soft X-ray background associated with the Galaxy known as the local hot bubble (LHB) and the Milky Way halo (MWH) in the FOVs of the {\em Suzaku} observations. In addition, the accurate estimate of the non X-ray background (NXB) in the {\em Suzaku} observations is important.

We first model the NXB of the {\em Suzaku} observations using the \texttt{xisnxbgen} task in HEAsoft \citep{Tawa2008}. The NXB spectra for our datasets are estimated from the dark-Earth data included in CALDB within a time interval of $\pm 150$ days around the observation date shown in Table~\ref{tbl:observation_logs}.

Next, we extract the X-ray spectra of the ICM from a region of $34'.5 - 40'.5$ toward the northwestern direction from the center to estimate the sky X-ray background. This region is shown as the red sector in Figure~\ref{fig:suzaku_image}, defined as the BGD region. The BGD region is located beyond $r_{200}$\footnote{$r_{200}$ represents a fiducial radius encompassing mean density equal to 200 times the critical density of the Universe at the redshift of the galaxy cluster.}, which is estimated approximately by $r_{200}=2.77~h_{70}^{-1}\,\mathrm{Mpc}~[kT/10\,\mathrm{keV}]^{1/2}/E(z)$, where $E(z)=[\Omega_\mathrm{M}(1+z)^3+1-\Omega_\mathrm{M}]^{1/2}$ \citep{Henry2009}. Thus, $r_{200}$ of A3667 is estimated to be 2.26\,Mpc $(= 34'.1)$ with a temperature of 7\,keV, based on the cosmology assumed in this paper and the cluster redshift of $z = 0.0556$.
      
We fit the X-ray spectra extracted from the BGD region with two types of spectral models (Case~1 and Case~2). Case~1 consists of three components of the sky X-ray background, i.e., LHB, MWH, and CXB, described as an unabsorbed thermal plasma, an absorbed thermal plasma, and a power-law component, respectively. Thus, Case~1 is expressed as $constant \times \{apec_\mathrm{LHB}+phabs\times(apec_\mathrm{MWH}+powerlaw_\mathrm{CXB})\}$, where the $\textit{apec}$ model in $\texttt{XSPEC}$ represents the ICM is in CIE \citep{Smith2001, Foster2012}. On the other hand, Case~2 includes another thermal plasma component with a temperature of 0.8\,keV, which is considered to be associated with diffuse X-ray emission in the Galaxy generated by, e.g., supernovae \citep[e.g.,][]{Yoshino2009, Sugiyama2023}. Thus, Case~2 is described as $constant \times \{apec_\mathrm{LHB}+phabs\times(apec_\mathrm{MWH}+powerlaw_\mathrm{CXB}+apec_\mathrm{0.8keV})\}$. In both cases, the redshift and abundance for all $\textit{apec}$ components are fixed at zero and 1$Z_{\odot}$, respectively. The photon index of a power-law model for CXB is fixed at 1.4 as determined by \citet{Kushino2002}. The model parameters for XIS1 are tied to those for the other XIS, except for the $constant$ parameter in both models. The $constant$ parameter for the XIS1 is fixed to 1, and the others are left free to vary. Although we adopt the energy range of $0.5-10\,\mathrm{keV}$ for the XIS0, XIS2, and XIS3, and $0.35-7.0\,\mathrm{keV}$ for the XIS1 in the spectral analysis, we mask the energy range around the Si-K edge ($1.82-1.84\,\mathrm{keV}$) to avoid systematic uncertainty from the response of the XIS \citep{Okazaki2018}. The X-ray spectra extracted from the XIS1 are fitted simultaneously with those from the other XIS. 

Figure~\ref{fig:result_bgd} shows the X-ray spectra extracted from the BGD region and their best-fit models derived from Case~1 and Case~2. The best-fit parameters are summarized in Table~\ref{tbl:bgd}. The obtained surface brightness of CXB in $2 - 10$\,keV is $7.55_{-0.31}^{+0.33}\times10^{-12}\,\mathrm{erg\,cm^{-2}\,s^{-1}}$ for Case~1 and $7.79_{-0.21}^{+0.22}\times10^{-12}\,\mathrm{erg\,cm^{-2}\,s^{-1}}$ for Case~2. The best-fit values of the temperature of LHB and MWH for Case~1 are $0.16\pm0.01\,\mathrm{keV}$ and $0.64\pm0.04\,\mathrm{keV}$, respectively, and those for Case~2 are $0.12_{-0.01}^{+0.02}\,\mathrm{keV}$ and $0.26_{-0.03}^{+0.04}\,\mathrm{keV}$, respectively. The fitting statistics for Case~1 and Case~2 show C-stat/d.o.f. of 4811.09/4003 and 4794.18/4402, respectively. 

We compare the best-fit results derived from different models using Akaike Information Criterion \citep[AIC:][]{Akaike1974} with the modified Cash statistics \citep{Cash1979}. The AIC value is calculated by $\mathrm{AIC}_i=2k_i-2\ln(L_i)$, where $k_i$ is the number of free parameters and $L_i$ is the maximum likelihood for model $i$. \cite{Cash1979} defines $C_i=-2\ln(L_i)$, where $C_i$ represents the Cash statistic value, i.e., $\mathrm{AIC}_i=2k_i+C_i$. In this paper, we measure goodness-of-fit for the spectra by calculating the difference in AIC values between models ($\Delta\mathrm{AIC}=\mathrm{AIC}_i-\mathrm{AIC_{min}}$), where the $\mathrm{{AIC}_{min}}$ is the minimum AIC value among all models. Since the likelihood $\exp(-\Delta\mathrm{AIC}/2)$ with AIC = 10 is approximately 0.0067, $\Delta\mathrm{AIC} > 10$ is commonly used as a threshold to distinguish between two models \citep[e.g.,][]{Burnham2004, Liddle2007, Arcodia2018, Benefield2019, Psaradaki2020, Ng2022}. Therefore, we adopt this threshold to evaluate the goodness-of-fit for the spectral.

The difference between the AIC values of Case~1 and Case~2, i.e., $\Delta \mathrm{AIC}$, is $18.91$ (i.e., $\Delta \mathrm{AIC} > 10$), indicating that Case~2 provides an improved fit to the observed spectra. Therefore, we adopt Case~2 for the sky X-ray background components and fix the best-fit parameters of Case~2 in the following spectral analysis.

\begin{deluxetable}{lcc}[htb]
   \tablecaption{Best-fit parameters of the X-ray spectral analysis for the sky X-ray background components} \label{tbl:bgd}
\tablewidth{0pt}

\tablehead{
   \colhead{Parameter} & \colhead{Case1} & \colhead{Case2}
}
\startdata
   $kT_\mathrm{LHB}\,\mathrm{(keV)}$ & $0.16\pm0.01$ & $0.12_{-0.01}^{+0.02}$ \\
   $N_\mathrm{LHB}$$~(\times10^{-3})$\tablenotemark{a} & $5.2\pm 0.6$ & $6.9_{-1.5}^{+1.9}$ \\
   $kT_\mathrm{MWH}\,\mathrm{(keV)}$ & $0.64\pm0.04$ & $0.26_{-0.03}^{+0.04}$ \\
   $N_\mathrm{MWH}$$~(\times10^{-3})$\tablenotemark{a} & $0.8\pm0.1$ & $1.7\pm0.5$ \\
   $\Gamma_\mathrm{CXB}$ & 1.4~(fixed) & 1.4~(fixed) \\
   $N_\mathrm{CXB}$$~(\times10^{-3})$\tablenotemark{b} & $1.18\pm0.04$ & $1.18\pm0.04$ \\
   $kT_\mathrm{0.8\,keV}\,\mathrm{(keV)}$ & $-$ & 0.8~(fixed) \\
   $N_\mathrm{0.8\,keV}$$~(\times10^{-4})$\tablenotemark{a} & $-$ & $4.2\pm1.0$ \\
   $\mathrm{C~statistic}/d.o.f.$ & 4811.09/4403 & 4794.18/4402 \\
   AIC & 13617.09 & 13598.18 \\
\enddata
\tablenotetext{a}{Normalization $N$ of the \textit{apec} component in XSPEC.}
\tablenotetext{b}{Units: $\mathrm{photons\,cm^{-2}\,s^{-1}\,keV^{-1}}$ at $1\,\mathrm{keV}$}
\end{deluxetable}

\subsection{X-ray Spectral Analysis between the Northwestern and the Southeastern relics}
\label{sec:center}

To measure the thermodynamic properties of the ICM in A3667, we define 11 source regions between the northwestern and the southeastern radio relics. These 11 source regions are from the region closest to the northwestern radio relic to the southeastern one across the cluster center, as shown in Figure~\ref{fig:suzaku_image}. The six regions in the northwestern direction from the center are separated by a radius of $4'.5, 9'.0, 13'.5, 18'.0, 22'.5, 27'.0$, and $31'.5$, respectively, within a sector with an opening angle of $36^{\circ}-81^{\circ}$ centered at (RA, Dec) $= (303^\circ.137, -56^\circ.829$). On the other hand, the four regions in the southeastern direction from the center are divided by a radius of $4'.5, 9'.0, 14'.0, 19'.0$, and $24'.0$, respectively, in a sector with an opening angle of $200^{\circ}-270^{\circ}$. The remaining region is defined to cover the central region of A3667 with a circle with a radius of $4'.5$. We extract and analyze the X-ray spectra of the ICM from all the defined regions.

\begin{deluxetable*}{lcccccc}[htb]
   \tablecaption{Best-fit parameters of the X-ray spectral analysis for A3667 derived from the CIE model and NEI model.} \label{tbl:cie_nei}
\tablewidth{0pt}
\tablehead{
   \colhead{Region (radius)} & \colhead{$kT\,\mathrm{(keV)}$} & \colhead{$Z~(\mathrm{Z_\odot})$} & \colhead{$n_et~(\mathrm{s\,cm^{-3}})$} & \colhead{$N~(\times10^{-3})$\tablenotemark{a}} & \colhead{$\mathrm{C~statistic}/\mathrm{d.o.f.}$} & AIC
}
\startdata
\multicolumn{1}{l}{CIE model:}\\
         1\tablenotemark{b} ~($-31'.5\sim-27'.0$)&$4.2_{-1.6}^{+3.5}$&$0.300$ (fixed)&$-$&$0.37_{-0.16}^{+0.30}$&$5085.96/4402$&13889.96\\
         2\tablenotemark{b} ~($-27'.0\sim-22'.5$)&$2.4_{-0.2}^{+0.3}$&$0.300$ (fixed)&$-$&$0.26\pm0.02$&$5050.60/4402$&13854.60\\
         3 ~~~($-22'.5\sim-18'.0$)&$7.0_{-1.2}^{+0.6}$&$0.180_{-0.107}^{+0.109}$&$-$&$2.82_{-0.45}^{+0.04}$&$4575.35/4401$&13377.35\\
         4 ~~~($-18'.0\sim-13'.5$)&$7.1\pm0.4$&$0.268_{-0.087}^{+0.094}$&$-$&$3.88_{-0.19}^{+0.09}$&$4798.97/4401$&13600.97\\
         5 ~~~($-13'.5\sim-9'.0$)&$7.5\pm0.5$&$0.259_{-0.098}^{+0.106}$&$-$&$2.53_{-0.10}^{+0.06}$&$4902.61/4401$&13704.61\\
         6 ~~~($-9'.0\sim-4'.5$)&$8.3\pm0.4$&$0.375_{-0.082}^{+0.095}$&$-$&$5.83_{-0.29}^{+0.09}$&$4627.32/4401$&13429.32\\
         7 ~~~($-4'.5\sim4'.5$)&$6.8_{-0.1}^{+0.2}$&$0.445_{-0.031}^{+0.028}$&$-$&$29.29_{-0.32}^{+0.19}$&$4610.21/4401$&13412.21\\
         8 ~~~($4'.5\sim 9'.0$)&$5.4\pm0.3$&$0.680_{-0.110}^{+0.120}$&$-$&$2.79_{-0.13}^{+0.08}$&$4760.46/4401$&13562.46\\
         9 ~~~($9'.0\sim14'.0$)&$6.0_{-0.4}^{+0.5}$&$0.267_{-0.093}^{+0.005}$&$-$&$1.89_{-0.13}^{+0.05}$&$4790.28/4401$&13592.28\\
         10\tablenotemark{b} ($14.'.0\sim19'.0$)&$5.7_{-0.9}^{+0.7}$&$0.300$ (fixed)&$-$&$1.14_{-0.16}^{+0.05}$&$4776.37/4402$&13580.37\\
         11\tablenotemark{b} ($19'.0\sim 24'.0$)&$3.2\pm0.6$&$0.300$ (fixed)&$-$&$0.46_{-0.06}^{+0.05}$&$5011.41/4402$&13815.41\\
   \hline
   \multicolumn{1}{l}{NEI model:}\\
        1\tablenotemark{b} ~($-31'.5\sim-27'.0$)&$4.0_{-1.4}^{+1.8}$&$0.300$ (fixed)&$2.0\times10^{13} ~(>1.0\times10^{10})$&$0.36_{-0.11}^{+0.31}$&$5085.88/4401$&13887.88\\
        2\tablenotemark{b} ~($-27'.0\sim-22'.5$)&$2.3_{-0.2}^{+0.3}$&$0.300$ (fixed)&$1.1\times10^{12} ~(>6.0\times10^{11})$&$0.25\pm 0.02$&$5050.14/4401$&13852.14\\
        3 ~~~($-22'.5\sim-18'.0$)&$6.9_{-1.7}^{+0.7}$&$0.180_{-0.101}^{+0.104}$&$4.6\times10^{13}~(>1.4\times10^{11})$&$2.81_{-0.62}^{+0.05}$&$4576.84/4400$&13376.84\\    
        4 ~~~($-18'.0\sim-13'.5$)&$6.8_{-0.4}^{+0.5}$&$0.255_{-0.085}^{+0.092}$&$4.3\times10^{13}~(>7.9\times10^{11})$&$3.88_{-0.19}^{+0.09}$&$4792.47/4400$&13592.47\\
        5 ~~~($-13'.5\sim-9'.0$)&$6.9_{-0.3}^{+0.5}$&$0.199_{-0.083}^{+0.102}$&$5.1\times10^{11}~(>2.9\times10^{11})$&$2.35_{-0.10}^{+0.05}$&$4881.91/4400$&13681.91\\
        6 ~~~($-9'.0\sim-4'.5$)&$8.3\pm0.4$&$0.362_{-0.141}^{+0.115}$&$1.8\times10^{12}~(>7.2\times10^{11})$&$5.73_{-0.25}^{+0.11}$&$4628.93/4400$&13428.93\\
        7 ~~~($-4'.5\sim-4'.5$)&$6.9_{-0.1}^{+0.2}$&$0.397_{-0.032}^{+0.042}$&$1.1_{-0.3}^{+2.6}\times10^{12}$&$29.37_{-0.25}^{+0.15}$&$4608.76/4400$&13408.76\\
        8 ~~~($4'.5\sim 9'.0$)&$6.6_{-0.6}^{+0.4}$&$0.351_{-0.070}^{+0.109}$&$2.2_{-0.4}^{+0.5}\times10^{11}$&$2.67_{-0.08}^{+0.09}$&$4723.78/4400$&13523.78\\
        9 ~~~($9'.0\sim14'.0$)&$5.9\pm0.5$&$0.254_{-0.098}^{+0.124}$&$6.1\times10^{11}~(>3.7\times10^{11})$&$1.89_{-0.14}^{+0.06}$&$4787.61/4400$&13587.61\\
        10\tablenotemark{b} ($14.'.0\sim19'.0$)&$6.2_{-0.6}^{+0.9}$&$0.300$ (fixed)&$4.7\times10^{13}~(>5.3\times10^{11})$&$1.16_{-0.25}^{+0.03}$&$4792.66/4401$&13594.66\\
        11\tablenotemark{b} ($19'.0\sim 24'.0$)&$3.0_{-0.5}^{+0.6}$&$0.300$ (fixed)&$2.1\times10^{12}~(>4.0\times10^{11})$&$0.43_{-0.05}^{+0.03}$&$5017.62/4401$&13819.62\\
\enddata
\tablenotetext{a}{Normalization $N$ of the \textit{apec} component 
 or \textit{nei} component in XSPEC.}
\tablenotetext{b}{The best-fit parameters of the spectra analysis for the CIE model or NEI model with the metal abundance at $0.3~Z_\odot$. The best-fit parameters when the metal abundance is set to be free are shown in Appendix~\ref{sec:appendix}}
\end{deluxetable*}

\subsubsection{Spectral fits with a CIE model}
\label{sec:fit_w_cie}

Here, we analyze the X-ray spectra of the ICM in A3667 using a CIE model to measure the thermodynamic properties of the ICM. We adopt an absorbed \texttt{apec} model (i.e., \texttt{phabs}$\times$\texttt{apec}) to describe the ICM X-ray emission. For the central source regions (Reg 3 to Reg 9), the temperature, metal abundance, and normalization of the model are set as free parameters. In contrast, for the outer regions (Reg 1, Reg 2, Reg 10, Reg 11), we fix the metal abundance of the model at 0.3\,$Z_\odot$, following the \cite{Sugawara2017}. 

Figure~\ref{fig:xis_all} shows the radial profiles of the thermodynamic properties of the ICM in A3667 between the northwestern and the southeastern relics, and the best-fit parameters of our spectral analysis are summarized in Table~\ref{tbl:cie_nei}. We compute the ICM electron number density from the spectral normalization $N$ of the $apec$ model, which is derived directly from \texttt{XSPEC} in the spectral analysis. The relation between $N$ and the ICM electron number density is given by
\begin{equation}\label{eq:norm}
   N=\frac{10^{-14}}{4\pi[D_\mathrm{A}(1+z)]^2}\int n_\mathrm{e}n_\mathrm{H}\mathrm{d}V,
\end{equation}
where $D_\mathrm{A}$ is the angular diameter distance, $V$ is the volume element, $n_\mathrm{e}$ and $n_\mathrm{H}$ are the electron and hydrogen density. Assuming $n_\mathrm{e}\sim 1.2n_\mathrm{H}$ and a line-of-sight (LOS) length of $L = 1$\,Mpc, we calculate the ICM electron number density\footnote{The volume of each region is calculated by multiplying a projected surface area of each region by the LOS length of $L = 1$\,Mpc.}. We also compute the ICM pressure and entropy as $P=n_\mathrm{e}\times kT$ and $K=kT \times n_\mathrm{e}^{-2/3}$, respectively, where $kT$ is the ICM temperature.

The observed temperature profile shows that the ICM within $19' ( = 1254\,\mathrm{kpc})$ from the cluster core appears isothermal with a temperature of $\sim 7$\,keV. This feature is consistent with that expected from merging clusters \citep[e.g.,][]{Kempner2003, Million2010}. In addition, the temperature jump is found in both the northwestern and the southeastern radio relics. We will discuss these features in Section~\ref{sec:revisiting}.

\subsubsection{Spectral fits with an NEI model}
\label{sec:fit_w_nei}

Here, we apply an NEI model to the ICM X-ray emission to investigate the plasma conditions in the ICM. As indicated by numerical simulations \citep[e.g.,][]{Akahori2010}, shocks induce deviations from the CIE condition in the ICM. In fact, \cite{Inoue2016} found that the ICM associated with strong shocks in A754 exhibits the NEI condition. Since A3667 has prominent double radio relics, A3667 is an ideal target to study the impact of mergers on the plasma conditions in the ICM.

\begin{figure*}[hbt]
   \begin{center}
      \includegraphics[width=16.8cm]{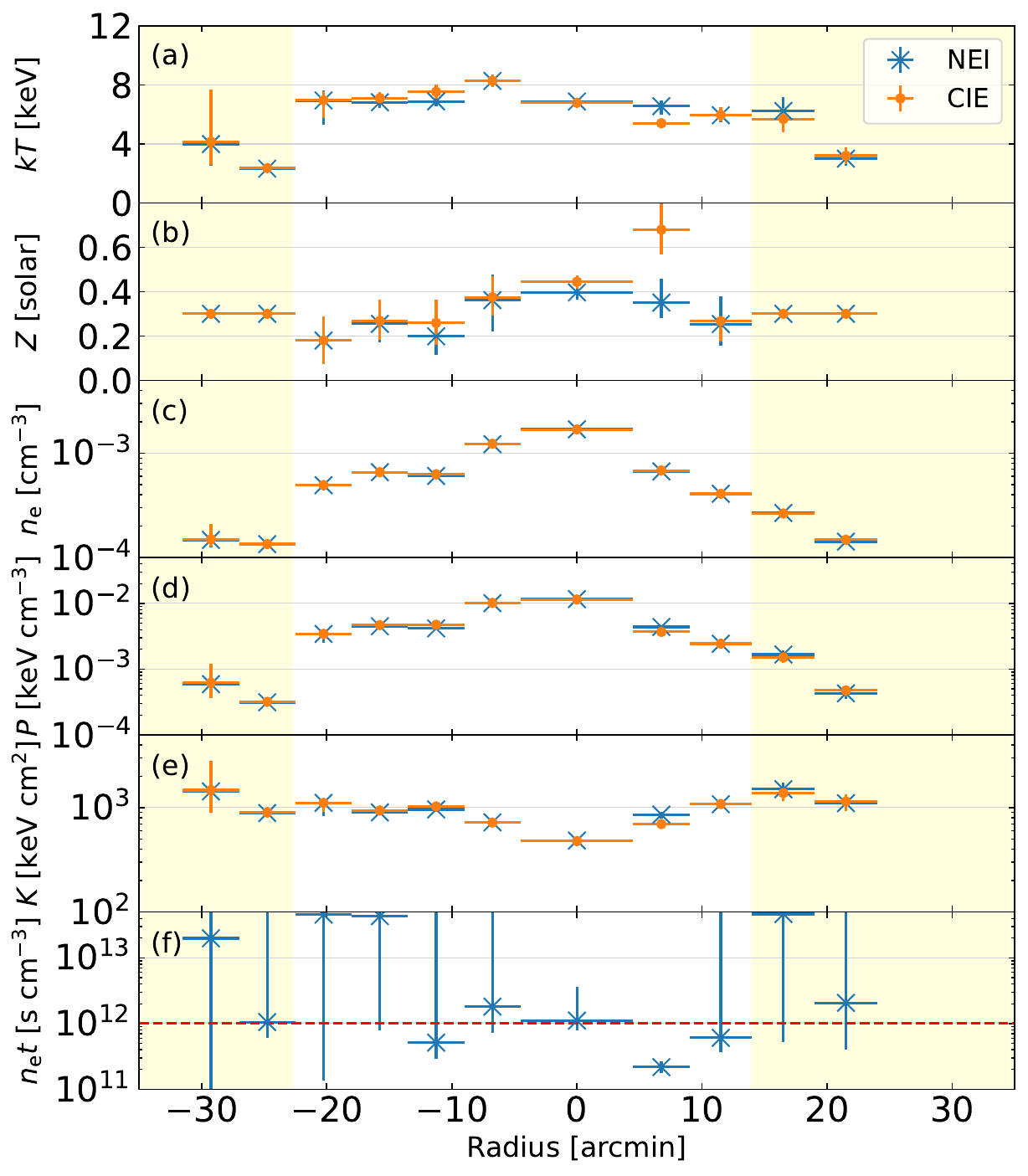}
   \end{center}
   \caption{
   Radial profiles of the thermodynamic properties of the ICM, including the ionization parameter, derived from the NEI model: From top to bottom, (a) the temperature $kT$, (b) the abundance $Z$, (c) the electron number density $n_\mathrm{e}$, (d) the pressure $P$, (e) the entropy $K$, and (f) the ionization parameter $n_\mathrm{e}t$. The orange and blue dots correspond to the best-fit results obtained with the CIE model and NEI model, respectively. The yellow-shaded areas represent the regions hosting the northwestern and southeastern radio relics. The horizontal axis corresponds to the distance from the center (RA, Dec) =  ($303^\circ.137, -56^\circ.829$) toward the southeastern direction. The distance toward the northwestern direction is defined as the negative value. In the bottom panel (f), the red dashed line shows the ionization parameter level of $n_\mathrm{e}t=1.0\times10^{12}\,\mathrm{s\,cm^{-3}}$ defined as the CIE state of the ICM.
   }\label{fig:xis_all}
\end{figure*}

We use an absorbed NEI model (i.e., \texttt{phabs}$\times$\texttt{nei}) to fit the observed X-ray spectra. Similar to the CIE model, we set the temperature and normalization to be free parameters in the fit. We also set the metal abundance of the ICM to be free, except in the outer regions. In addition, the ionization parameter $n_\mathrm{e}t$ is left free to vary in the NEI model. The obtained radial profiles of the ICM properties are shown in Figure~\ref{fig:xis_all}, and the best-fit parameters are summarized in Table~\ref{tbl:cie_nei}. 

We find that the best-fit ionization parameter for Reg~8 is measured to be $2.2_{-0.4}^{+0.5}\times10^{11}$\,s\,cm$^{-3}$, which is lower than that expected from the CIE state of the ICM (i.e., ${n_{\rm e}t} = 10^{12}$\,s\,cm$^{-3}$). Since the value of $\Delta\mathrm{AIC}$ between the CIE model and the NEI model is 38.68, the NEI model provides a better fit. In the case of Reg~7, its best-fit ionization parameter is $1.1_{-0.3}^{+2.6}\times10^{12}$\,s\,cm$^{-3}$, which is consistent with the CIE state of the ICM. In fact, the value of $\Delta\mathrm{AIC}$ between the CIE model and NEI model is $3.45$, indicating that the NEI model does not provide a better fit, in contrast to Reg~8. Interestingly, Reg~8 includes the prominent cold front in A3667. On the other hand, the ICM located in the regions of the radio relics (i.e., Reg~1, Reg~2, Reg~10, and Reg~11) seems consistent with the CIE state. We obtain only the lower limit of the ionization parameter in these regions.

\begin{figure}[htb]
   \begin{center}
      \includegraphics[width=8.5cm, trim={0.7cm 0.2cm 0.7cm 0cm}]{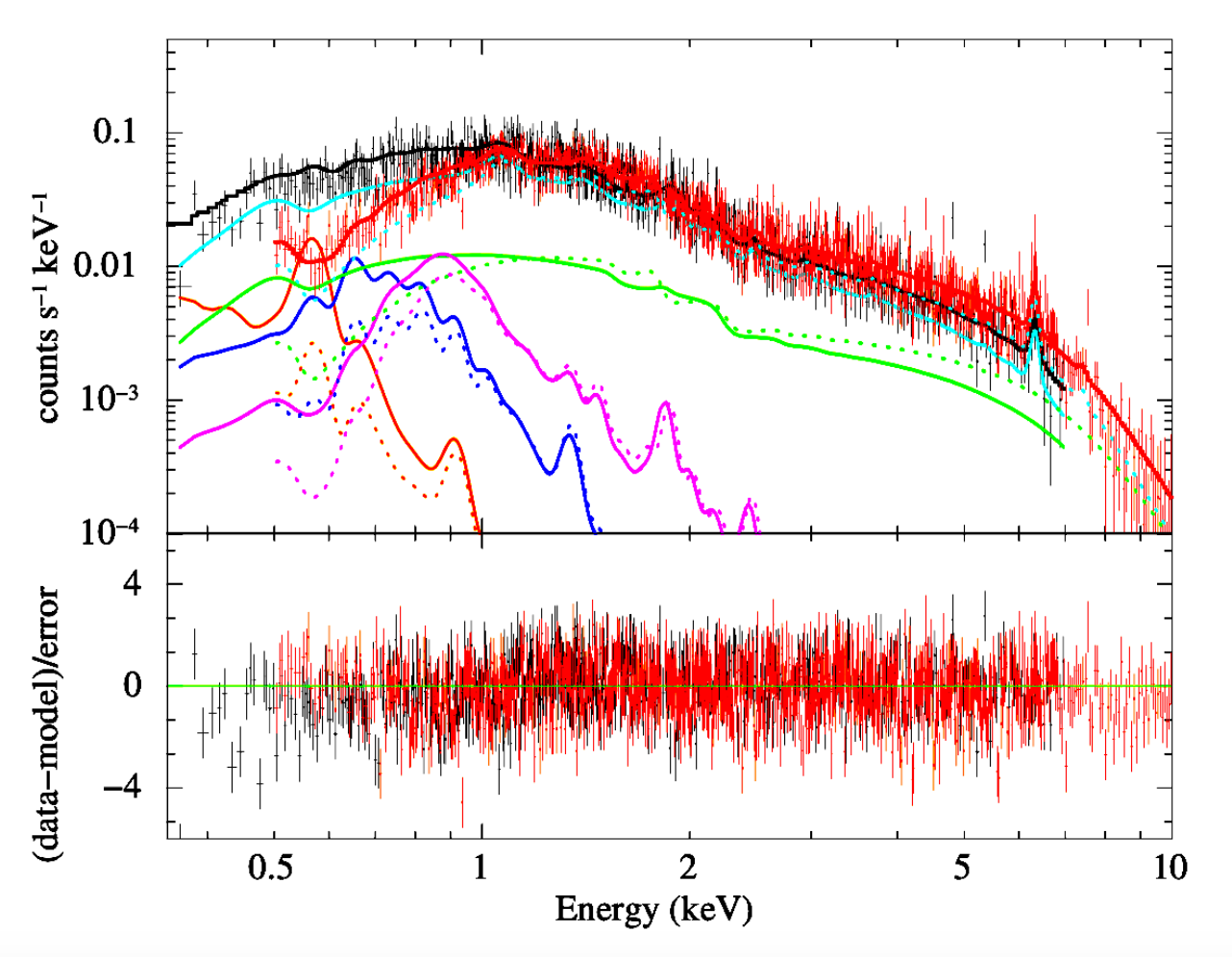}
   \end{center}
   \caption{
   Same as Figure~\ref{fig:result_bgd}, but for the observed X-ray spectra extracted from Reg~8 shown in Figure~\ref{fig:suzaku_image}. The Case~2 model is used for the sky X-ray background components. Assuming the NEI model, the X-ray emission from the ICM is shown in light blue. The black and the red solid lines represent the best-fit models for XIS1 and XIS0, XIS2, and XIS3, respectively. 
   }\label{fig:reg8_nei_case2}
\end{figure}

\subsection{X-ray Spectral Analysis across the Cold Front}
\label{sec:cf}

A3667 is known to host the prominent cold front in the southeastern direction, which is located within Reg~8. In Section~\ref{sec:fit_w_nei}, we have found that the ICM exhibits the NEI state in Reg~8. Here,
we investigate where the location of the NEI state of the ICM in Reg~8 is. To do so, we divide Reg~8 into two sub-regions, i.e., the inner and the outer regions defined as the inside and the outside of the cold front, respectively, using the {\em Chandra} X-ray image of A3667.

\begin{figure}[htb]
   \begin{center}
      \includegraphics[width=8.85cm, trim={3.9cm 9cm 3.5cm 7.8cm},clip]{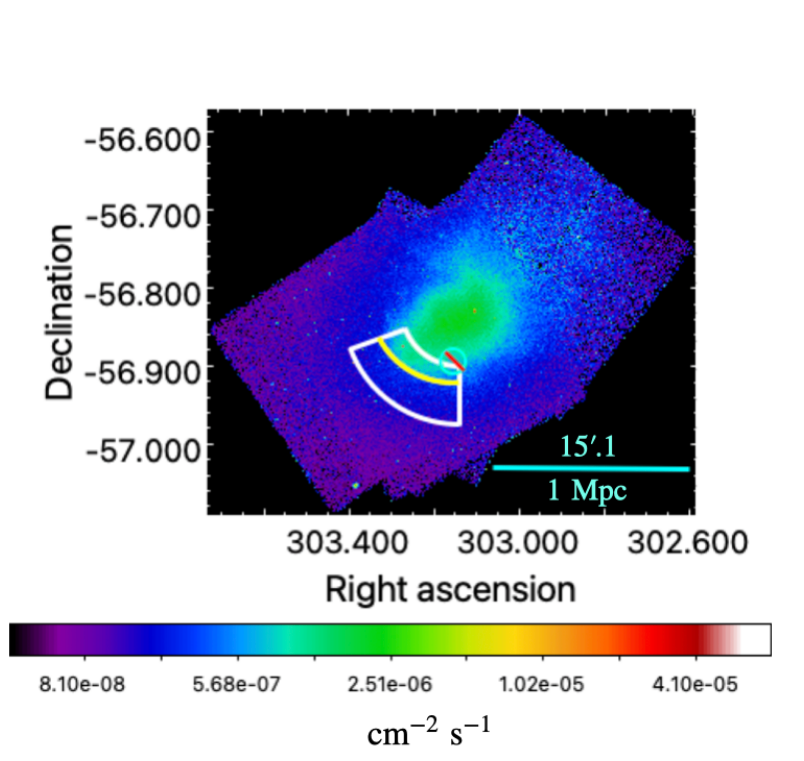}
   \end{center}
   \caption{X-ray surface brightness image of A3667 observed with {\em Chandra} in the $0.7-7.0\,\mathrm{keV}$ band. The yellow arc shows the position of the interface of the cold front. In the white sector, the inner and the outer regions are defined as the regions below and above the interface, respectively. The cyan circles show the point sources detected with {\em XMM-Newton}.  
   }\label{fig:chandra_image}
\end{figure}

We create the {\em Chandra} X-ray image of A3667 with the archival datasets mentioned in Table~\ref{tbl:observation_logs} to determine the position of the interface of the cold front. We reprocess all the datasets using the $\tt{chandra\_repro}$ task in {\em Chandra} Interactive Analysis of Observations \citep[CIAO;][]{Fruscione2006} version 4.16.0 with $\tt{CALDB}$ 4.11.1. We use the $\tt{merge\_obs}$ task in \texttt{CIAO} to create a mosaic X-ray image of all the {\em Chandra} observations in the $0.5-7.0\,\mathrm{keV}$ band, and exposure map created at 2.3\,keV. The exposure-corrected {\em Chandra} image of A3667 is shown in Figure~\ref{fig:chandra_image}. 
We measure the position of the boundary of the cold front to be located $6'.4$ within a sector spanning $210^{\circ}-270^{\circ}$ from the center at (RA, Dec) $= (303^\circ.152, -56^\circ.817)$. This position is used as a boundary between the inner and the outer regions (see Figure~\ref{fig:chandra_image}).

\begin{figure}[htb]
   \begin{center}
      \includegraphics[width=8.5cm, trim={0.4cm 0.2cm 0.4cm 0cm}]{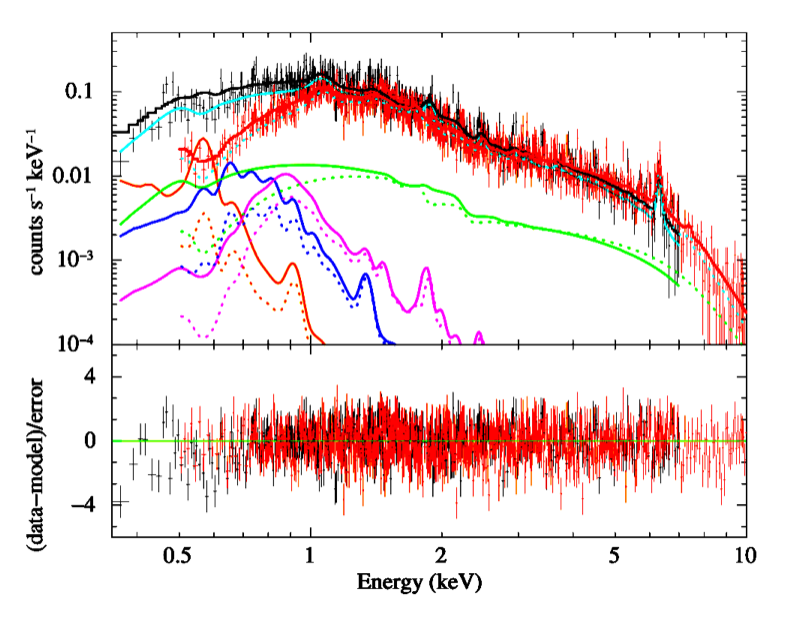}
   \end{center}
   \caption{
   Same as Figure~\ref{fig:reg8_nei_case2}, but for the observed X-ray spectra extracted from the inner region shown in Figure~\ref{fig:chandra_image} and their best-fit models.
   }\label{fig:inner_nei_case2}
\end{figure}

We fit the X-ray spectra of the ICM extracted from the inner and the outer regions in the same way as Section~\ref{sec:center}, i.e., a CIE model or an NEI model. The free parameters of these two models are treated in the same way as our previous analysis. The difference from our previous analysis is that the value of the redshift is set to be free, considering possible gas motions around the cold front. The observed X-ray spectra of the ICM in the inner region are shown in Figure~\ref{fig:inner_nei_case2}, and the best-fit parameters derived from the CIE and the NEI models are summarized in Table~\ref{tbl:1kt_coldfront}. 

Interestingly, we find that the ICM in the inner region exhibits the NEI state with an ionization parameter of $n_\mathrm{e}t=(2.5\pm0.7)\times10^{11}\,\mathrm{s\,cm^{-3}}$, which is consistent with that for Reg~8. The value of $\Delta$AIC between the CIE and the NEI models is $15.91$, indicating that the NEI model offers a better fit. On the other hand, the ICM in the outer region exhibits no NEI state. We obtain the lower limit of the ionization parameter in the outer region. In fact, the value of $\Delta$AIC between the CIE and the NEI models is $1.43$, meaning that the NEI model does not improve the fit. Therefore, these results indicate that the NEI state of the ICM is located in the inner region only. 

\begin{deluxetable}{lcc}[htb]
   \tablecaption{Best-fit Parameters for the inner and the outer regions using the CIE model, the NEI models, and the CIE+CIE model..} \label{tbl:1kt_coldfront}
\tablewidth{0pt}
\tablehead{
   \colhead{Region} & \colhead{Inner} & \colhead{Outer}
}
\startdata
\multicolumn{1}{l}{CIE model:}
    \\$kT_\mathrm{CIE}\,\mathrm{(keV)}$&$5.0_{-0.2}^{+0.4}$&$5.9_{-0.4}^{+1.3}$\\
   $\mathrm{N}_\mathrm{CIE}$$~(\times10^{-3})\tablenotemark{a}$&$4.9_{-0.2}^{+0.4}$&$2.3_{-0.2}^{+0.4}$\\
   $Z~(\mathrm{Z_\odot})$&$0.8_{-0.1}^{+0.2}$&$0.4\pm0.1$\\
   Redshift&$0.058_{-0.001}^{+0.003}$&$0.055\pm0.005$\\
   $\mathrm{C~statistic}/d.o.f.$&$4347.42/4400$&$4783.13/4400$\\
   AIC&13147.42&13583.13\\
   \hline
   \multicolumn{1}{l}{NEI model:}\\
   $kT_\mathrm{NEI}\,\mathrm{(keV)}$&$6.0\pm0.7$&$5.7_{-0.4}^{+0.8}$\\
   $n_et~(\times 10^{11}~\mathrm{s\cdot cm^{-3}})$&$2.5\pm0.7$&$4.8 ~(>2.7)$\\
   $\mathrm{N}_\mathrm{NEI}$$~(\times10^{-3})\tablenotemark{b}$&$4.8_{-0.4}^{+0.1}$&$2.3\pm0.4$\\
   $Z~(\mathrm{Z_\odot})$&$0.5\pm0.1$&$0.3_{-0.1}^{+0.2}$\\
   Redshift&$0.059_{-0.002}^{+0.003}$&$0.055_{-0.004}^{+0.006}$\\
   $\mathrm{C~statistic}/d.o.f.$&$4333.51/4399$&$4786.56/4399$\\
   AIC&13131.51&13584.56\\
   \hline
   \multicolumn{1}{l}{CIE+CIE model:}\\
   $kT_\mathrm{CIE_{low}}\,\mathrm{(keV)}$&$4.1_{-0.2}^{+0.4}$&$-$\\
   $\mathrm{N}_\mathrm{CIE_{low}}(\times10^{-3})\tablenotemark{a}$&$3.7_{-3.2}^{+1.0}$&$-$\\
   $kT_\mathrm{CIE_{high}}\,\mathrm{(keV)}$&$19.2~(>9.6)$&$-$\\
   $\mathrm{N}_\mathrm{CIE_{high}}(\times10^{-3})\tablenotemark{a}$&$1.2_{-0.8}^{+2.2}$&$-$\\
   $Z~(\mathrm{Z_\odot})$&$0.9_{-0.1}^{+0.2}$&$-$\\
   Redshift&$0.057\pm0.003$&$-$\\
   $\mathrm{C~statistic}/d.o.f.$&4328.76/4398&$-$\\
   AIC&13124.76&$-$\\
\enddata
\tablenotetext{a}{Normalization $N$ of the \textit{apec} component in XSPEC.}
\tablenotetext{b}{Normalization $N$ of the \textit{nei} component in XSPEC.}
\end{deluxetable}

\section{Discussion}
\label{sec:discussion}

\subsection{NEI states of the ICM}

We have found that the ICM in the inner region shows an ionization parameter of $n_\mathrm{e}t=(2.5\pm 0.7)\times 10^{11}\,\mathrm{s\,cm^{-3}}$, indicating that the ICM located inside the cold front in A3667 is in an NEI state.
In general, the plasma of the ICM is expected to be in a CIE state. The timescale of thermal equilibration for CIE ($t_{\rm eq}$) is expressed as
\begin{equation}\label{equation:time_eq}
t_{\rm eq} \sim 7 \times 10^{7} \bigg( \frac{n_{\rm e}}{10^{-3}\,{\rm cm}^{-3}} \bigg)^{-1} \bigg(\frac{kT}{4\,{\rm keV}} \bigg)^{3/2} \,{\rm yr}, \label{eq:teq}
\end{equation}
\citep[e.g.,][]{Spitzer1962, Zeldovich1966}.
Since $t_{\rm eq}$ is much shorter than the expected age of galaxy clusters ($\sim 10^{10}$\,yr), the X-ray emission from the ICM is well-modeled using a CIE model. However, the equilibration timescale for electrons is much shorter than $t_{\rm eq}$ \citep[i.e., $\sim 10^{4}$\,yr, e.g., see][]{Spitzer1962}. Therefore, the NEI state of the ICM is possible within a short timescale after heating e.g., by shocks induced by cluster mergers. In fact, numerical simulations have predicted the presence of the NEI state of the ICM associated with shocks \citep[e.g.,][]{Takizawa1999, Akahori2010}. In addition, observationally, \cite{Inoue2016} have found the NEI state of the ICM in the shock front in a major merging cluster A754.

To compare the timescale derived from the ionization parameter $n_\mathrm{e}t$ with the thermal equilibration timescale for CIE~($t_{\rm eq}$), we estimate both the timescales for the inner region. The electron number density $n_\mathrm{e}$ of the inner region is calculated to be $1.76_{-0.07}^{+0.02}\times10^{-3}\>\mathrm{cm^{-3}}$ by Equation~\ref{eq:norm}. Applying this electron number density $n_\mathrm{e}$ to the measured ionization parameter, the timescale is inferred to be $4.6\pm1.2\,\mathrm{Myr}$. On the other hand, the thermal equilibration timescale for CIE in the inner region is calculated to be $74\pm12\,\mathrm{Myr}$ by Equation~\ref{equation:time_eq}. The timescale calculated from the ionization parameter is much shorter than the thermal equilibration timescale. This suggests that the ICM of the inner region has not yet reached a CIE state.

\subsubsection{NEI states and the cold front in A3667}

Our results indicate that the ICM is in the NEI state only in the inner region, i.e., inside the cold front in A3667. Given that the ICM in the radio relic regions is consistent with the CIE state, it is difficult to directly associate the NEI state of the ICM inside the cold front with shocks typically associated with radio relics. Therefore, we discuss here the presence of the ICM in the NEI state associated with the gas dynamics inside the cold front in A3667.

Two mechanisms are proposed for explaining the origin of cold fronts: the stripping scenario and the sloshing scenario \citep{Markevitch2007}. Stripping cold fronts are expected to be found in the cluster outskirts located in front of an infalling subcluster and to be formed by ram-pressure \citep[e.g.,][]{Markevitch2000, Vikhlinin2001a}. On the other hand, sloshing cold fronts are considered to be found in the cluster center and to be generated by gas sloshing resulting from gravitational disturbance triggered by an infalling subcluster \citep[e.g.,][]{Tittley2005, Ascasibar2006, Roediger2011}.
 
The cold front in A3667 has been widely considered to be a stripping cold front \citep[e.g.,][]{Vikhlinin2001, Ichinohe2017}. However, there is an alternative possibility that the cold front in A3667 is due to gas sloshing, namely a sloshing cold front. If the A3667 has a sloshing cold front, the weak shocks may induce the ICM in the NEI state inside the cold front in A3667.  In the case of a sloshing cold front, transonic motions of the gas are expected to be induced by gas sloshing, leading to the generation of weak shocks. Such weak shocks could change the plasma conditions of the ICM, leading the ICM to an NEI state. In fact, the cold fronts in and A2319 have been considered a sloshing one, and the sloshing cold front hosts transonic motions of the gas located at the head of the sloshing cold front \citep{Ueda2024}. Therefore, if the hypothesis that the gas sloshing induces the transonic motions associated with the cold front in A3667 is correct, the presence of the NEI state of the ICM might be able to be explained by its weak shocks. 

We proposed the hypothesis that the ICM in the NEI state inside the cold front in A3667 is associated with a weak shock inside the cold front. However, it is difficult to arrive at a firm conclusion regarding these hypotheses due to the lack of studies on the plasma conditions in the ICM. Further studies are needed to understand the NEI state of the ICM better. The X-ray microcalorimeter, Resolve, onboard {\em XRISM} \citep{Tashiro2018} is expected to be able to diagnose the plasma conditions in the ICM, thanks to its superb high energy resolution \citep{XST2020}. Therefore, {\em XRISM} observations of merging clusters will provide us with a unique opportunity to look into the ionization states of the ICM inside the cold front in A3667.

\subsubsection{Possible relations of the NEI states to radio halos}

A3667 is known to host the double giant radio relics in the northwestern and the southeastern directions \citep{Rottgering1997, Carretti2013, Hindson2014, Riseley2015, Gasperin2022}. Recently, \cite{Gasperin2022} have identified an elongated radio halo bridging the region between the northwestern radio relic and the prominent cold front in A3667 across the cluster center with MeerKAT, which was previously pointed out by \cite{Carretti2013}. Interestingly, no diffuse radio emission is observed between the outside of the cold front and the southeastern radio relic \cite{Gasperin2022}.

The NEI state of the ICM in A3667 seems to be associated with the elongated radio halo. In particular, our results show that the ICM exhibits the NEI state in the inner region, i.e., inside the cold front, suggesting that the NEI state may coincide with the edge of the elongated radio halo. One of the possible interpretations is that this elongated radio halo may be generated by turbulence in the wake of the infalling subcluster forming the prominent cold front \citep{Gasperin2022}. If this hypothesis is the case, the NEI state of the ICM might be related to such turbulence.

The NEI state of the ICM in A754 is also found at the edge of an elongated radio halo. \cite{Macario2011} presented the elongated radio halo in A754 observed with GMRT. Interestingly, the edge of the elongated radio halo coincides with the location of the ICM that shows the NEI state \citep{Inoue2016}, similar to our findings in A3667.

The coincidence between the NEI state of the ICM and the edge of the radio halo found in A3667 and A754 indicates that the NEI state of the ICM seems to prefer to be in radio halos rather than in radio relics. We have obtained the lower limit of the ionization parameter of the ICM located in the radio relics, which means that the plasma state in the ICM in the radio relic region is consistent with the CIE state (see Figure~\ref{fig:xis_all}). Although radio relics are considered not only to be a good tracer for shock fronts \citep[e.g.,][]{Ensslin1997, Akamatsu2013} but also to be used to explore the microphysical properties of the ICM, radio halos may be a good indicator to search for the NEI state of the ICM.

\subsection{Two-Temperature CIE Models for the NEI region}
\label{sec:2T}

We have found the NEI state of the ICM in the inner region, which is below the prominent cold front in A3667 (see Section~\ref{sec:cf}). However, as studied by \cite{Inoue2016}, two-temperature CIE models may be applicable to explain the observed X-ray spectra instead of a single-temperature NEI model. Therefore, we investigate this hypothesis using a CIE+CIE model (\texttt{phabs}$\times$(\texttt{apec}+\texttt{apec})). In this analysis, the metal abundance and the redshift of the two models are tied to each other. The best-fit parameters for the inner and the outer regions derived from the CIE+CIE model are summarized in Table~\ref{tbl:1kt_coldfront}.

We find that a very high-temperature CIE component is required for the inner region along with a $\sim 4$\,keV CIE component, while no secondary CIE component is required for the outer region. For the inner region, we obtain the lower limit of the temperature of the secondary CIE component to be $>9.6$\,keV. Interestingly, \citet{Nakazawa2009} reported the presence of a very hot thermal emission with a temperature of $19.2_{-4.0}^{+4.7}\,\mathrm{keV}$ using the Hard X-ray Detector (HXD) onboard {\em Suzaku}. Since the HXD is a non-imaging hard X-ray instrument \citep{Kokubun2007}, it is difficult to determine the emission region of such a very hot gas. In addition, the FOV of the HXD covers the elongated radio halo as well as part of the southeastern radio relic. Thus, there are several possibilities for interpreting the origin of the very hot thermal emission detected by the HXD. The observed normalization of the secondary CIE component is a factor of $30$ fainter than that for the very hot thermal emission reported by \citet{Nakazawa2009}. In addition, the metal abundance in the inner region becomes larger under the assumption of the CIE+CIE model compared to that in the cluster center (see Figure~\ref{fig:xis_all}). 
We also calculate the difference in $\Delta\mathrm{AIC}$ between the CIE+CIE model and the single-temperature NEI model in the inner region, and then we obtain $\Delta\mathrm{AIC}=6.75$. Therefore, although the possibility of the very hot thermal emission is not ruled out, the NEI state hypothesis may be preferred to interpret the observed results.

\subsection{Revisiting the global features of the thermodynamic properties of the ICM in A3667}
\label{sec:revisiting}

The entropy of the ICM in the cluster center is measured at $484_{-10}^{+12}\,\mathrm{keV\,cm^2}$, which is significantly higher than that observed in cool cores of galaxy clusters \citep[e.g.,][]{Cavagnolo2009, Sanderson2009, Hudson2010}. In contrast, the observed entropy profile agrees with that expected from merging clusters. Even if A3667 originally hosted a cool core, strong merger activities that are ongoing in A3667 may destroy a cool core.

The temperature of the ICM within $19'$ ($=1254$\,kpc) across the center of A3667 is about 7\,keV, which is consistent with the mean temperatures reported by \cite{Markevitch1998}, \cite{Vikhlinin2001a}, and \cite{Reiprich2002}. However, the ICM temperature profile shows a jump in the vicinity of the relics (i.e., Reg2 to Reg3: $-27'.0$ to $-18.0$, Reg10 to Reg11: $14'.0$ to $24'.0$), as shown in Figure~\ref{fig:xis_all}. If these temperature jumps are due to shocks, the Mach number $M$ can be estimated using the Rankine-Yugoniot jump condition as
\begin{equation}
    \frac{T_2}{T_1}=\frac{5M^4+14M^2-3}{16M^2},
\end{equation}
where $T_1$ is a pre-shock temperature of Reg2 or Reg11 and $T_2$ is a post-shock temperature of Reg3 or Reg10. The values of the inferred Mach number are $M=2.6_{-0.4}^{+0.3}$ at the northwestern relic and $M=1.8_{-0.4}^{+0.3}$ at the southeastern relic. 
The Mach number around the southeastern relic is consistent with the previous reports \citep[e.g.,][]{Storm2018, Akamatsu2013}. Additionally, around the northwestern relic, our result is consistent with that reported in \cite{Akamatsu2013} and \cite{Sarazin2016}.

\section{Conclusions}
\label{sec:conclusions}

In this paper, we have analyzed the thermodynamic properties of the intracluster medium (ICM), including the ionization parameter as an indicator of the presence of the non-equilibrium ionization (NEI) state in the merging cluster A3667 between the northwestern and the southeastern relics using {\em Suzaku}. 
The main conclusions of this paper are as follows:

\begin{enumerate}

\item  We observe that the ICM inside the cold front in A3667 indicates an NEI state with an ionization parameter $n_\mathrm{e}t$ of $(2.5\pm0.7)\times 10^{11}\,\mathrm{s\, cm^{-3}}$ at the $90\%$ confidence level. This value is lower than the $n_\mathrm{e}t > 10^{12}\,\mathrm{s\, cm^{-3}}$ expected in a collisional ionization equilibrium (CIE) state. The value of $\Delta\mathrm{AIC}$ between the CIE and NEI models is 15.91 (i.e., $\Delta\mathrm{AIC} >10$), indicating that the NEI model provides a better fit to the X-ray spectra of the ICM extracted from inside the cold front. In addition, the timescale derived from this ionization parameter is $4.6\pm1.2\,\mathrm{Myr}$, which is significantly shorter than the thermal equilibration timescale. This suggests that the ICM inside the cold front has not yet reached a CIE state. 

\item The presence of the ICM in the NEI state inside the cold front in A3667 might be explained by a weak transonic sloshing motion inside the cold front. If A3667 has a sloshing cold front, there is a possibility that a weak shock induced by the transonic motion of the gas could explain the NEI state observed in the ICM. In support of this hypothesis, \citet{Ueda2024} indicated that A2319, which is considered to have a sloshing cold front, exhibits transonic gas motions at the head of the front. Therefore, in the event that the gas sloshing in A3667 similarly induces transonic motions, the NEI state of the ICM could be attributed to these weak shocks.

\item The NEI state of the ICM in A3667 appears to be associated with the elongated radio halo. The NEI state of the ICM in A3667 is observed inside the cold front. This result suggests that the NEI state may coincide with the edge of the elongated radio halo. A possible explanation is that this radio halo might be induced by turbulence in the wake of the infalling subcluster that forms the prominent cold front \citep{Gasperin2022}. If this hypothesis is correct, the NEI state of the ICM might be related to such turbulence. 

\item The radio halos might be a good indicator to search for the NEI state of the ICM. Interestingly, the NEI state is also found by \cite{Inoue2016} in the ICM at the edge of an elongated radio halo of A754, similar to our results in A3667. In addition, we obtain only the lower limit of the ionization parameters $n_\mathrm{e}t$ in the regions located in the radio relics, which means that the ICM in the radio relics is consistent with the CIE state. These results suggest that the NEI state of the ICM seems to prefer to be in radio halos rather than in radio relics.

\item We measured the entropy of the ICM in the cluster center at $484_{-10}^{+12}\,\mathrm{keV\,cm^2}$, which is significantly higher than that observed in cool cores of galaxy clusters. In contrast, the observed entropy profile agrees with that expected from merging clusters. Even if there was originally a cool core in A3667, the strong fusion activity that has taken place in A3667 can destroy a cool core. 

\item The ICM temperature profile appears to be isothermal with a mean temperature of around 7\,keV within $19' (=1254$\,kpc) from the center of A3667. We also observed temperature jumps and estimated the Mach numbers $M$ of the shocks associated with the northwestern and the southeastern relics to be $2.6_{-0.4}^{+0.3}$ and $1.8_{-0.4}^{+0.3}$, respectively. Our results are in good agreement with those derived from the previous reports.

\end{enumerate}

\section{Acknowledgments}
We deeply appreciate Shutaro Ueda for the discussions and for carefully proofreading the manuscript. We also thank Hiroki Akamatsu for providing helpful comments and the MeerKAT data of A3667. The results presented in this work are made possible by the successful efforts of the {\em Suzaku} and {\em Chandra} teams, who have contributed to the observatory's construction, launch, and operation. We thank the ASIAA Summer Student Program 2024 for its hospitality and for allowing us to discuss the research. This work was supported by JSPS KAKENHI Grant Number 22H01269 and 21K13946.

\vspace{5mm}
\facilities{Suzaku, CXO}

\software{astropy \citep{AstropyCollaboration2013,Collaboration2022}, CIAO \citep{Fruscione2006},  
          HEASoft \citep{NHEASARC2014},
          SAOImage DS9 \citep{SAO2000},
          XSPEC \citep{Arnaud1996}
          }

\appendix
\section{Uncertainty of the metal abundance in the outskirts of A3667 to measure the plasma conditions of the ICM}

In Section~\ref{sec:ana}, we have fixed the metal abundances of the ICM at 0.3\,$Z_\odot$ for Reg~1, Reg~2, Reg~10, and Reg~11 in the spectral analysis to measure the thermodynamic properties of the ICM in these regions. Here, under this assumption, we investigate systematic uncertainties of the ICM properties, including the plasma conditions. To this end, we set the metal abundance as a free parameter in the spectral analysis using both the CIE and the NEI models. 

As shown in Table~\ref{tbl:cie_nei_free}, the best-fit values of the metal abundance are lower than $0.3~Z_\odot$ for all regions we analyzed here. However, the best-fit values of the other parameters are consistent with those measured in Section~\ref{sec:ana}. Therefore, we conclude that assuming the metal abundance of 0.3\,$Z_\odot$ for the ICM in the outskirts does not affect the measurements of the plasma conditions of the ICM.

\label{sec:appendix}
\begin{deluxetable*}{lcccccc}[htb]
   \tablecaption{Best-fit Parameters for the outskirts the CIE model and NEI model.\label{tbl:cie_nei_free}} \label{tbl:bgd_free}
\tablewidth{0pt}
\tablehead{
   \colhead{Region (radius)} & \colhead{$kT\,\mathrm{(keV)}$} & \colhead{$Z~(\mathrm{Z_\odot})$} & \colhead{$n_et~(\mathrm{s\,cm^{-3}})$} & \colhead{$N~(\times10^{-3})$\tablenotemark{a}} & \colhead{$\mathrm{C~statistic}/\mathrm{d.o.f.}$} & AIC
}
\startdata
   \multicolumn{1}{l}{CIE model:}\\
        1 ~($-31'.5\sim-27'.0$)&$3.8_{-1.7}^{+4.5}$&$0.093~(<0.094)$&$-$&$0.38_{-0.11}^{+0.33}$&$5084.02/4401$&13886.02\\
        2 ~($-27'.0\sim-22'.5$)&$1.9_{-0.3}^{+0.4}$&$4.933\times 10^{-7}~(<0.079)$&$-$&$0.35_{-0.05}^{+0.04}$&$5037.27/4401$&13839.27\\
        10 ($14.'.0\sim19'.0$)&$5.7_{-1.0}^{+0.7}$&$0.063~(<0.218)$&$-$&$1.19_{-0.16}^{+0.06}$&$4770.65/4401$&13572.65\\
        11 ($19'.0\sim 24'.0$)&$2.8_{-0.8}^{+0.6}$&$1.376\times 10^{-12}~(<0.092)$&$-$&$0.54_{-0.06}^{+0.08}$&$5001.92/4401$&13803.92\\
   \hline
   \multicolumn{1}{l}{NEI model:}\\
        1 ~($-31'.5\sim-27'.0$)&$3.8_{-2.0}^{+4.2}$&$0.079~(<0.340)$&$4.7\times10^{13}~(>1.0\times10^{10})$&$0.38_{-0.11}^{+0.33}$&$5083.92/4400$&13883.92\\
        2 ~($-27'.0\sim-22'.5$)&$1.8_{-0.2}^{+0.4}$&$9.067\times 10^{-14}~(<0.068)$&$5.5\times10^{11} ~(>1.6\times10^{11})$&$0.35_{-0.06}^{+0.04}$&$5036.06/4400$&13836.06\\
        10 ($14.'.0\sim19'.0$)&$6.4_{-0.8}^{+0.7}$&$0.034~(<0.143)$&$3.0\times10^{11}~(>1.3\times10^{11})$&$1.19_{-0.11}^{+0.07}$&$4786.84/4400$&13586.84\\
        11 ($19'.0\sim 24'.0$)&$4.0_{-1.0}^{+0.8}$&$1.044\times10^{-18}~(<0.822)$&$5.8\times10^{11}~(>1.1\times10^{11})$&$0.61\pm0.07$&$5007.44/4400$&13807.44\\
\enddata
\tablenotetext{a}{Normalization $N$ of the \textit{apec} component 
 or \textit{nei} component in XSPEC.}
\end{deluxetable*}

\bibliography{ref}{}
\bibliographystyle{aasjournal}

\end{document}